# Combination prevention for the elimination of HIV†

Brian G. Williams

South African Centre for Epidemiological Modelling and Analysis (SACEMA), Stellenbosch, South Africa

Correspondence to BrianGerardWilliams@gmail.com

## Abstract

The development of potent drugs for the control of viraemia in people living with HIV means that infected people may live a normal, healthy life and offers the prospect of eliminating HIV transmission in the short term and HIV infection in the long term. Other interventions, including the use of condoms, pre-exposure prophylaxis, treatment of sexually transmitted infections and behaviour change programmes, may also be effective in reducing HIV transmission to varying degrees.

Here we examine recommendations for when to start treatment with anti-retroviral drugs, estimate the impact that treatment may have on HIV transmission in the short and in the long term, and compare the impact and cost of treatment with that of other methods of control. We focus on generalized HIV epidemics in sub-Saharan Africa. We show that universal access to ART combined with early treatment is the most effective and, in the long term, the most cost-effective intervention. Elimination will require effective coverage of about 80% or more but treatment is effective and cost effective even at low levels of coverage.

Other interventions may provide important support to a programme of early treatment in particular groups. Condoms provide protection for both men and women and should be readily available whenever they are needed. Medical male circumcision will provide a degree of immediate protection for men and microbicides will do the same for women. Behaviour change programmes in themselves are unlikely to have a significant impact on overall transmission but may play a critical role in supporting early treatment through helping to avoid stigma and discrimination, ensuring the acceptability of testing and early treatment as well as compliance.

## Introduction

Treatment guidelines for people living with HIV increasingly advise people to start anti-retroviral therapy (ART) sooner rather than later after infection. However, treatment guidelines have changed significantly over the past fifteen years and there are differences among country guidelines leading to uncertainty as to how to interpret and act on the advice.

Here we examine historical changes in the treatment guidelines and argue that $CD4^+$ cell counts should no longer be used as a condition for starting anti-retroviral therapy (ART). We develop a simple model to determine the impact that each of several interventions could have on HIV-transmission. We investigate the impact of early treatment, use of condoms, pre-exposure prophylaxis, treatment of sexually transmitted infections and behaviour change programmes on the incidence of HIV if given to individual people and at a population level in the short and the long term.

We show that the elimination of HIV, using currently available methods of control, will depend primarily on early treatment but other methods of control will provide additional support and will provide an important degree of protection for some groups at high risk of infection. Finally, we examine the relative impact and costs of the various interventions and briefly discuss the affordability of stopping HIV.

## Changing guidelines

Several international organizations, including the International AIDS Society (IAS), the European AIDS Clinical Society (EACS), the World Health Organization (WHO), and many countries[1] have developed guidelines on when to start ART. The guidelines have changed over time and vary among countries.[2] A particularly important change was made in 2002 when the WHO advised low- and middle-income countries to start adults on ART only when their $CD4^+$ cell count was below 200/μL.

In the year 2000 the International AIDS Society (IAS) and the Department of Health and Human Services (DHHS) used data from the Multi-Centre AIDS Cohort Study (MACS)[3] (Figure 1) to assess the probability that a person infected with HIV would develop an AIDS related condition in the next three years.

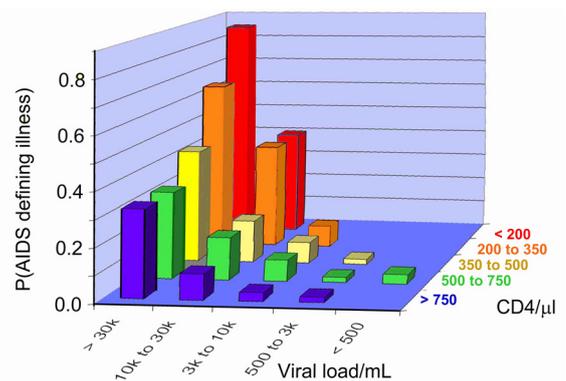

Figure 1. Probability of developing an AIDS related condition within 3 years as a function of viral load and $CD4^+$ cell count[4] from the MACS cohort.[3]

As expected, people with low viral loads are unlikely to develop AIDS related conditions especially if they also have a high $CD4^+$ cell count. The data in Figure 1 are replotted in Figure 2 as an area plot. The inset numbers give the probability of developing an AIDS related condition. For example, people whose viral load and $CD4^+$ cell counts puts them on the line dividing the green and yellow areas have a 5% probability of developing an AIDS related condition in the next three years. The brown dots in Figure 2 are from a cross-sectional survey of HIV-positive young men in Orange Farm, South Africa in the year 2000 (Bertran Auvert, personal communication) and show the range of viral load and $CD4^+$ cell counts in a typical community in South Africa.

---

† This paper is a development of a presentation delivered at the 3rd International HIV Treatment as Prevention (TasP) Workshop held in Vancouver, BC, Canada, April 21 to 25, 2013 and in part at the 7th International Workshop on HIV Treatment, Pathogenesis and Prevention Research in Resource-limited Settings, held in Dakar, Senegal, May 14 to 17, 2013.

In Figure 2, top right, the DHHS recommendations for starting ART[4] in the year 2000 are indicated by the blue lines. Everyone with a viral load above 10k/mL, to the right of the blue lines, or with a $CD4^+$ cell count below 500/µL, below the blue lines, were advised to start ART. The brown dots show that in Orange Farm, in the year 2000, 90% of HIV-positive young men would have been eligible for ART under the DHHS guidelines.

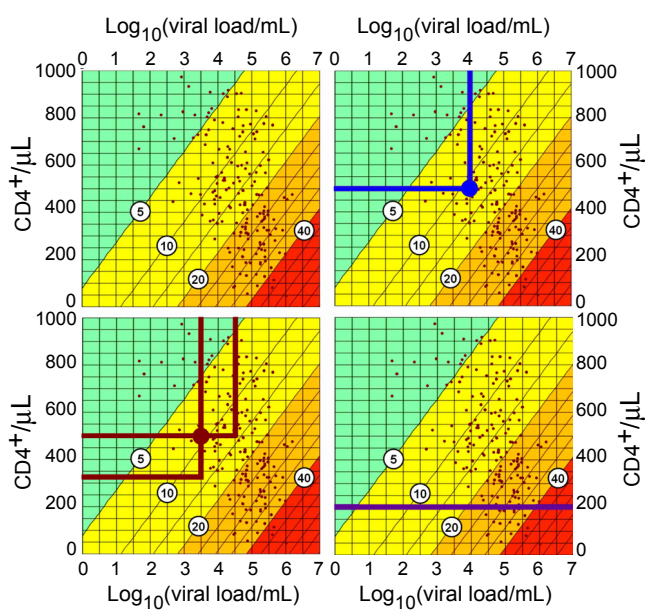

Figure 2. Top left: The data in Figure 1 replotted as an area plot. The brown dots are viral load and $CD4^+$ cell counts for young men in Orange Farm, South Africa in the year 2000 (Bertran Auvert, personal communication). Top right: DHHS 2000 guidelines;[4] people to the right of and below the blue lines should start ART. Bottom left: IAS 2000 guidelines;[5] people to the right of and below the brown lines should start ART; those between the brown lines should consider starting ART. Bottom right: WHO 2002 guidelines;[6] people below the purple line should start ART. Inset numbers: probability (%) of developing an AIDS related condition in the next three years.

In Figure 2, bottom left, the IAS recommendations for starting ART[5] also in the year 2000 are indicated by the brown lines. Everyone with a viral load above 30k/mL, to the right of the brown lines, or with a $CD4^+$ cell count below 350/µL, below the brown lines, was advised to start ART. Everyone with a viral load and $CD4^+$ cell count that puts them between the lines should consider starting therapy. The brown dots show that in Orange Farm, in the year 2000, 90% of HIV-positive young men would have been eligible for ART under the IAS guidelines.

In Figure 2, bottom right, the WHO recommendations for starting ART, in the year 2002, are indicated by the purple line.[6] They do not include a viral load cut-off and recommend only that people with a $CD4^+$ cell count below 200/µL were advised to start ART. The brown dots show that in Orange Farm, in the year 2000, 10% of HIV-positive young men would have been eligible for ART under the WHO guidelines. It is important to note that the WHO guidelines also recommend treatment for all people in WHO clinical stages III or IV. We can estimate the number of people who will develop an AIDS related opportunistic infection before they reach a $CD4^+$ cell count of 200/µL from data on the incidence of various opportunistic infections as a function of $CD4^+$ cell counts.[7] These data suggest that approximately 20% of people infected with HIV will develop AIDS related conditions including *Pneumocystis carinii* pneumonia, *Mycobacterium avium* complex infection, toxoplasmosis, cytomegalovirus, or a fungal infection before their $CD4^+$ cell count falls to 200/µL. Furthermore, about half of those who develop TB will do so before their $CD4^+$ cell count reaches 200/µL.[8]

The effect of the changes introduced by WHO in 2002 was to reduce the proportion of HIV-positive young men in Orange Farm who would have been eligible for treatment from 90% to 10% whereas one might have expected a 'public health approach' to be concerned with ways to expand, not reduce, treatment coverage.

Many countries continue to follow the WHO 2002 guidelines and in 2012 about 30% of countries surveyed recommended ART only for those asymptomatic people with a $CD4^+$ cell count below 200/µL and about another 60% of countries surveyed recommended ART only for those asymptomatic people with a $CD4^+$ cell count below 350/µL.[1]

After the year 2002 most of the formal guidelines have steadily increased the $CD4^+$ cell count at which people are advised to start treatment. In 2013 the IAS[9] recommended that 'All adults with HIV infection should be offered ART regardless of CD4 cell count, based on …data [showing] that all patients may benefit from ART …[and] that ART reduces the likelihood of HIV transmission [and] provides clinical benefits'. In 2013 the DHHS[10] recommended 'ART … for all HIV-infected individuals to reduce the risk of disease progression… [and]… for the prevention of transmission'. In both sets of guidelines it was felt that early ART was in the best interests of the individual patient and would have the added benefit of preventing further transmission.

Now, in 2013, the WHO recommends treatment for all people whose $CD4^+$ cell counts is less than 500/µL and all HIV-positive people with TB, Hepatitis B, who are pregnant, or under the age of five years, irrespective of $CD4^+$ cell count. The new guidelines are close to where they were in the year 2000 but without the viral load condition.[11] Data on the distribution of $CD4^+$ cell counts in HIV-negative people[12] suggest that about 80% of all those currently infected with HIV and not on ART will have a $CD4^+$ cell count below 500/µL. If we include the other groups of people that should start treatment irrespective of their $CD4^+$ cell count then about 90% of all HIV positive people are currently eligible for ART. Since $CD4^+$ cell counts have very little prognostic value at an individual level,[12] considerable savings in time, human resources and money could be had by abandoning the use of $CD4^+$ cell counts for deciding on when to start ART.

If there is a need to triage people in order to treat those at greatest risk first, the sensible way to do this would be on the basis of each person's viral load. Individual $CD4^+$ cell counts can vary by an order of magnitude within populations,[12] the mean $CD4^+$ cell count can vary by a factor of two between populations,[12] and survival is independent of the initial $CD4^+$ cell count.[12,13] $CD4^+$ cell counts therefore have very little prognostic value except in that unfortunate circumstance when the count is very low by which time an infected person is likely to be in WHO clinical stages III or IV and in need of immediate treatment anyway. People with a high viral load, on the other hand, have a greatly reduced life expectancy[14] and are much more infectious than those with a low viral load.[15] Where the availability of anti-retroviral drugs is limited, giving preference to people with high viral loads would have the greatest benefit for individual patients and the greatest impact on transmission.[16]

## Ending the epidemic

Given that early ART is in the best interests of individual patients the question arises as to role of early treatment in ending the epidemic of HIV. Other ways of reducing transmission have been studied including condom promotion (CP), pre-exposure prophylaxis (PreP), medical male circumcision (MC), treatment of other sexually transmitted infections (STI), and behaviour change programmes (BC). Each of these could be used singly or in combination and we consider the contribution that each of them might make to ending the epidemic.

Table 1. Incidence rate ratios (*IRR*) for *treatment-as-prevention* (TasP), condom promotion (CP), pre-exposure prophylaxis (PreP), medical male circumcision (MC), treatment of sexually transmitted infections (STI), and behaviour change programmes (BC). Trial conditions: *IRR* values obtained from trials; Coverage: *IRR* allowing for estimated coverage given in Figure 3B; Long term: impact on long term steady state incidence allowing for non-linearities and heterogeneity assuming a steady state prevalence of 16% and $R_0 = 4.5$ (Appendix 2). Exp.: expected values; 'Low' and 'High': 95% confidence limits.

|  | Trial conditions | | | Coverage | | | Long-term | | |
|---|---|---|---|---|---|---|---|---|---|
|  | Exp. | Low | High | Exp. | Low | High | Exp. | Low | High |
| TasP | 0.003 | 0.000 | 0.022 | 0.202 | 0.200 | 0.218 | 0.208 | 0.197 | 0.300 |
| CP | 0.128 | 0.040 | 0.407 | 0.564 | 0.520 | 0.704 | 0.837 | 0.806 | 0.911 |
| MC | 0.400 | 0.310 | 0.490 | 0.816 | 0.778 | 0.850 | 0.953 | 0.940 | 0.963 |
| PreP | 0.430 | 0.321 | 0.575 | 0.772 | 0.729 | 0.830 | 0.938 | 0.922 | 0.957 |
| STI | 0.883 | 0.762 | 1.024 | 0.907 | 0.810 | 1.019 | 0.978 | 0.951 | 1.004 |
| BC | 0.948 | 0.810 | 1.111 | 0.959 | 0.848 | 1.089 | 0.991 | 0.962 | 1.017 |

TasP: based on a randomized controlled trial (RCT),[17] a meta-analysis[18] and data on viral load and transmission.[15] CP: comparing 'always' v. 'never' used condoms in 12 studies;[19] PreP: four RCTs[20-23] excluding those stopped for futility;[24,25] MC: three RCTs;[26-28] STI: three community RCTs;[29-31] BC: ten community RCTs.[32]

## Projecting the impact

We first establish the magnitude of the control problem, that is to say the degree to which transmission should be reduced in order to reduce the case reproduction number, $R_0$, to less than 1.[33]

For countries in sub-Saharan Africa the value of $R_0$ can be estimated from the initial doubling time of the prevalence of HIV and the life expectancy of people infected with HIV but not on ART. This gives a median value of 4.5 (80% within 2.6–6.3; range 1.6–9.5).[34] To eliminate HIV in half the countries of sub-Saharan Africa transmission must be reduced by 4.5 times, or by 78%, and to eliminate HIV in 90% of the countries transmission must be reduced by 84%.

## The impact of interventions

Trials of the different interventions have been done and the results are summarized in Table 1 and shown in Figure 3A. The result for TasP is an estimate based on the relationship between transmission and viral load[15] and is consistent with the results of the HPTN 052 trial[17] and a meta-analysis.[18] The result for condom use is an average across 12 groups of people who said that they always uses condoms compared to those that said that they never used condoms.[19] The result for PreP is from four randomized controlled trials (RCTs)[20-23] but excluding the Fem-PreP[24] and Voice trials[25] which were stopped for reasons of futility, probably because they were unable to achieve sufficiently high levels of compliance.[35] The result for the impact of behaviour change interventions is based on ten community-RCTs.[32]

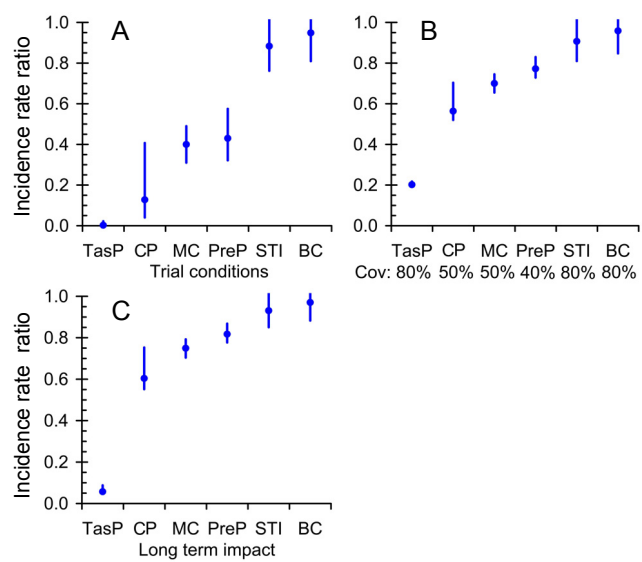

Figure 3. The incidence rate ratio for: people on TasP; people who 'always' use condoms (CP); for men who are circumcised (MC); people on PreP; for the treatment of other sexually transmitted infections (STI); and for behaviour change programmes. A: results from trials (see text for further details); B: allowing for the effective coverage levels indicated; C: the reduction in the long-term incidence in South Africa with $R_0 = 5.8$ and the current adult prevalence of 16%.

For the treatment of sexually transmitted infections the impact is measured in the whole population among which only some people have a sexually transmitted infection. If the reduction in the individual level of transmission is $\rho$ and the prevalence of STIs is $P$, then the average reduction, $r$, would be $r = \rho P + (1-P)$. We can calculate the reduction in the individual level of transmission if one person is treated from the reduction in risk in the whole population if we know the prevalence of curable STIs.[36] The estimated value of $\rho$ is given in Figure 3A. In the case of behaviour change interventions we assume that the individual risk reduction is equal to the overall level of risk reduction.

To assess the short and long term population level impact and cost of the different interventions we discount the impact of the trials to allow for the effective coverage, including both coverage and compliance, as shown in Figure 3B. Here we assume that under TasP one can achieve an effective coverage of 80%; this allows for a testing coverage of 85% and annual testing, which should eliminate 90% of transmission in those tested. In the case of condom promotion we assume, perhaps optimistically, that people can be persuaded to use condoms in half of all sexual encounters. In the case of male circumcision some men will already be circumcised while others may refuse to be circumcised. Since the focus of PrEP would be on people at especially high risk we assume that about 40% of sexual encounters are covered by PrEP; in concentrated epidemics UNAIDS estimates that less than half of all transmission events are in key populations at high risk of infection. We assume that the effective circumcision coverage may reach 50% of all sexually active men and we allow for the fact that male circumcision protects men but not women.[37] For sexually transmitted infections we assume that treatment reduces the prevalence of STIs by 80%; the reason why the impact is small even if those with an STI are at very high risk of infection, is that typically only about 10%[36] of adults may have a sexually transmitted infection at a given time. In the case of behaviour change programmes we assume that coverage is 80%.

Finally we wish to estimate the long term impact of the various interventions assuming that the epidemic reaches a new steady state allowing for non-linearities in the relationship between transmission and prevalence and for heterogeneity in sexual activity (Appendix 1). In Figure 3C we use the estimated increase in $\alpha$, the rate at which people start treatment under TasP, and the estimated declines in $\lambda$, the reduction in the risk of infection for the other interventions, to estimate the expected steady-state incidence that would be achieved if each of the various interventions was implemented according to the effective coverage assumptions in Figure 3B. Figure 3C then shows the importance of the non-linearity in the impact of any intervention as a function of the reduction in transmission.

## Economics of control

Of the order of US$16 billion per year is currently being spent on understanding and dealing with the epidemic of HIV/AIDS in low- and middle-income countries.[38] Expanding treatment and prevention efforts will demand significant investments, both financial and social, and we need to consider the economics of control. We first consider the cost-effectiveness of the available interventions and then refer briefly to the question of affordability since all interventions must be paid for.

### Cost

We assume that the costs associated with each of the interventions are as follows.
1. TasP: $250 per person per year for drugs, $250 per person per year for support and $5 per test (Appendix 2).
2. PrEP: $1 per day and $5 per monthly test;[39,40]
3. MC: $50 per circumcision;[41,42]
4. STI: $20 per treatment;[43]
5. BC: $0.3 per person reached per year.[44]

Effectiveness can be measured in many ways including direct disease outcomes such as prevalence, incidence, mortality or life expectancy, but also social and economic outcomes such as community coherence, reduction in stigma, increases in employment, contribution to the gross national product and so on. Since we are concerned with controlling and eventually eliminating HIV we will focus on the impact on HIV incidence. We do not include the considerable cost-savings that will accrue through avoiding hospitalization, medical costs for people who would otherwise have developed AIDS-related conditions or the economic benefits of maintaining a healthy and productive workforce. This analysis is conservative in relation to costs and benefits.

In estimating the costs and impact of intervening with a single person we assume that the background incidence and prevalence are unchanged and use the reduction in the risk of infection as measured in the trials (Figure 3A and Table 1).

For the short term, which we may consider to be the first year after full implementation, we make the same assumptions but allow for estimated levels of coverage (Figure 3B and Table 1).

In the long term, say ten years after reaching full coverage, we assume that the epidemic is in a new steady state. The point is that people infected with HIV must either be started on ART or they will die. Since the financial cost to society of letting them die is, in almost all countries, greater than the cost of maintaining them on ART (Appendix 3), we assume that under interventions other than TasP HIV-positive people are started on ART at a late stage of their disease so that they are kept alive and are no longer infectious but the intervention has no impact on transmission. We must then account for the cost of keeping all HIV-positive people alive once they start ART in all of the interventions.

### Costs and effects

Details of the cost-benefit calculations are given in Appendix 4 and the impact of each intervention in relation to its cost is shown in Figure 4 for each intervention in South Africa assuming an initial, steady state prevalence among adults of 16% and a value of $R_0$ equal to 5.8. The shaded areas indicate the cost-benefits ratios. For example, the line dividing the areas shaded green and blue would imply a cost of $100 per adult per year to reduce the

incidence to zero, or a cost of $50 per adult per year to reduce the incidence by 50%.

If individual people are started on ART the cost and the impact on an individual person's risk of infection is shown in Figure 4A. The most effective interventions are TasP and using condoms since these have the greatest impact at an individual level. The most cost effective intervention is treating sexually transmitted infections in those that present with such an infection but this only applies to the relatively small proportion of people who have a sexually transmitted infection at a given time. PreP is the third most effective but is the least cost effective. For PreP we assume that people are taking or using anti-retroviral drugs every day even though they may only be occasionally at risk. If a microbicide could be applied only when the person is expecting to have sex, then the costs could be cut by an order of magnitude. This would reduce the cost of PreP to something like the cost of using condoms or male circumcision. Behaviour change programmes are very cost effective but the effect is very small so they are unlikely to have an impact.

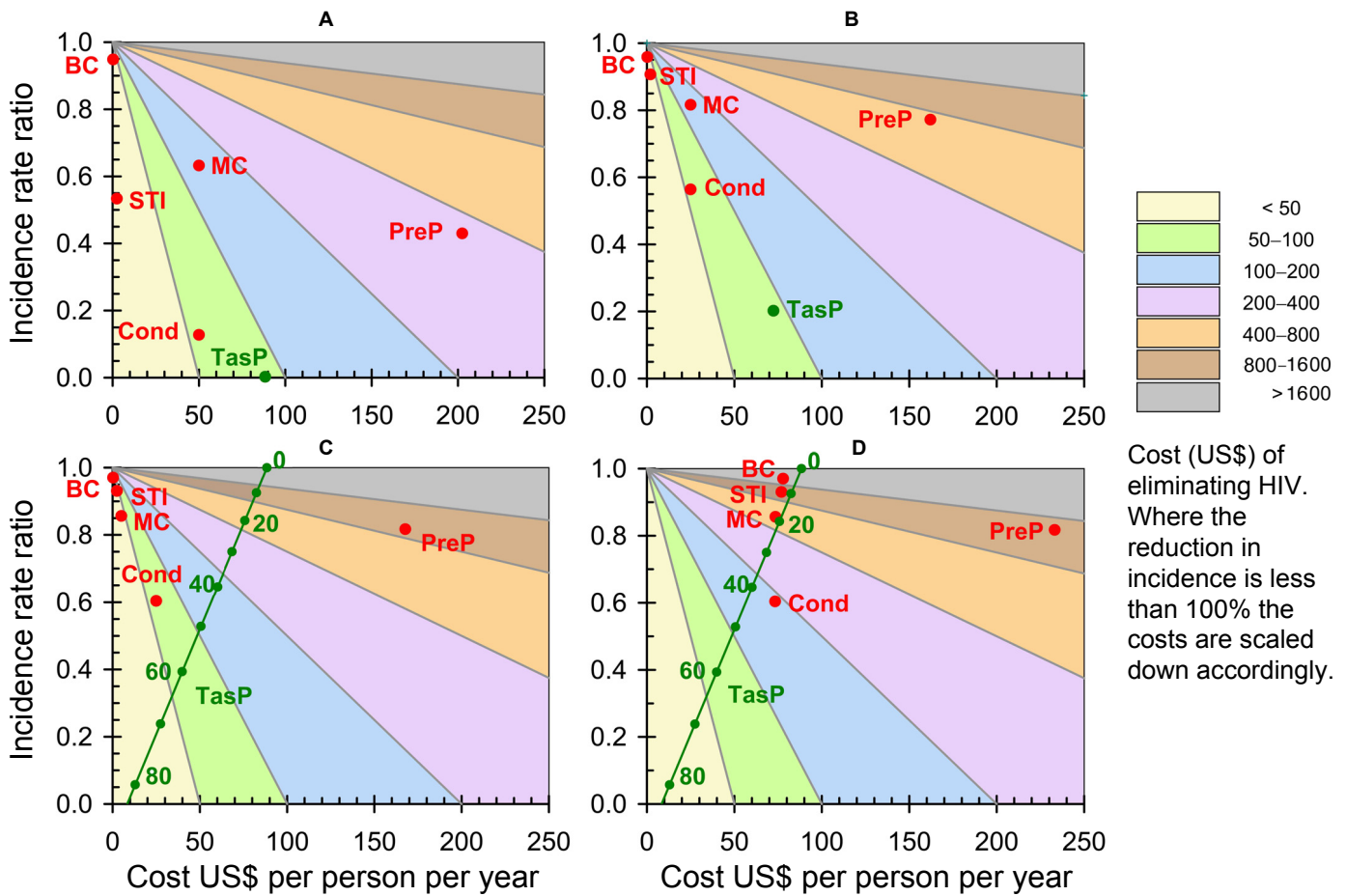

Figure 4. The HIV incidence rate ratio, with and without each intervention, plotted against the cost of various interventions in South Africa assuming an initial steady state prevalence of 16% and a value of $R_0$ equal to 5.8. The incidence rate ratio is normalized to the initial incidence so that 1 corresponds to an incidence of 1.6% per year. The shaded areas indicate benefits per unit cost for fixed cost-benefit ratios (see legend). A: the costs and benefits of making the interventions available to individual people; B: the short term costs and benefits of each intervention assuming the coverage rates indicated in Figure 3; C: the long term costs and benefits of each intervention. D: as C but including the costs of maintaining people on ART for life assuming they start treatment just before they would otherwise have died. MC: male circumcision; BC: behaviour change; PreP; Pre-exposure prophylaxis: CP: condom promotion; STI: treating sexually transmitted infections; TasP: Treatment-as-prevention. In C and D the numbered dots indicate percentage levels of effective coverage under TasP.

Considering the impact of each intervention at a population level, but only in the short term which we consider to be the first year after full implementation, we need to allow for the levels of coverage and compliance, discussed in the text and indicated in Figure 3. For male circumcision we also have to allow for the fact that it protects men but not women. In this case the costs and the benefits are reduced in all cases but the impact per unit cost remains more or less the same (Figure 4B). TasP, at 80% coverage and compliance, still has the greatest impact and the impact per unit cost is among the best while PreP remains the least cost-effective but subject to the caveats noted in the previous paragraph concerning the frequency of use.

Considering the long term costs and benefits of each intervention we allow for the non-linear relationship between reductions in transmission and reductions in the steady state prevalence (Appendix 1). We start by excluding the cost of letting people die without ART (Figure 4C). Since TasP alone has the potential to end the epidemic of HIV if $R_0$ is reduced below the critical value of 1, the long term costs and benefits for TasP depend critically on the levels of coverage and compliance that can be reached. In Figure 4C we plot the costs and benefit for effective coverage of TasP ranging from 0 to 80% (green dots in steps of 10%). TasP still has the greatest potential impact on incidence provided the coverage is greater than about 50%.

The major long term cost arises from the need to maintain those that are infected with HIV on ART for life remembering that letting them die will cost even more. In Figure 4D we plot the results in Figure 4C but including the long term costs of maintaining all HIV-positive people on ART assuming that they start ART just before they would otherwise have died. The cost of TasP is unchanged since it is assumed that everyone is maintained on ART for life. The cost of PreP is much the same because the major cost is the provision of drugs to HIV-negative people. TasP at about 40% effective coverage has the same impact and cost-benefit as condom promotion, at 20% has the same impact as male circumcision, at 10% has the same impact as treating sexually transmitted diseases and at 5% has the same impact as behaviour change programmes.

## Discussion

### Ethics of early treatment

In their 2002 guidelines for starting ART the WHO noted that 'the resolve of the international community to address this appalling disparity between treated and untreated, between rich and poor, is stronger than ever. The world recognizes the pressing moral, social, political and economic need to expand access to antiretroviral therapy to many more millions of people living with HIV/AIDS as soon as practicable. … Countries are encouraged to adopt a public health approach in order to facilitate the scale-up of ARV use in resource-limited settings. This means that antiretroviral treatment programmes should be developed and that ARV treatment should be standardized'. The 2002 WHO guidelines continue: 'The recommendation to start treatment in asymptomatic patients only when the CD4 count drops below 200/μL takes account of the following major unanswered question relating to ART: when should treatment be initiated in the setting of established infection among asymptomatic HIV-positive persons? While beginning therapy before the CD4 cell count falls below 200/μL clearly provides clinical benefits, the actual point above 200/μL at which to start therapy has not been definitively determined.' One might have argued that since 'beginning therapy before the CD4 cell count falls below 200/μL clearly provides clinical benefits' one should rather have recommended starting earlier and if evidence that starting later became available, one could then have recommended starting later. It is unfortunate that adapting a public health approach led to considerably fewer people being eligible to start treatment.

Given the current recommendations of the IAS and the DHHS, both based on extensive and detailed analyses of the risks and benefits to individual patients as well as the likely impact on transmission, it is clear that early ART will keep people alive, stop them from developing AIDS related infections, stop them from infecting their partners and has the potential to end the epidemic. Taking into account the most recent recommendations of the WHO[11] it would be unethical and in violation of the Hippocratic Oath to deny ART to anyone infected with HIV on the basis of their $CD4^+$ cell count.

### Affordability

In addition to the costs and the cost-effectiveness it is important to consider the affordability of eliminating HIV. This depends on the size of the epidemic in each country, the wealth of that country and the overall cost if it proves to be necessary to receive donor funding. In a previous study we therefore compared the cost of maintaining every HIV-positive person in each country in sub-Saharan Africa on ART with the gross domestic product (GDP) of that country.[45] The country that is least able to afford universal ART is Malawi where the cost would amount to just less than 8% of GDP and there are four other countries for which the cost would exceed 5% of GDP: Lesotho, Zimbabwe, Mozambique and Burundi. Interestingly, Malawi already has one of the best ART programmes in the world but these five countries will clearly need international assistance to deal with the epidemic. However, the total cost for all five countries would only amount to US$1.4 billion so that the international community could easily afford to provide the necessary support. In a further six countries, Uganda, the Central African Republic, Tanzania, Zambia, Swaziland and Kenya the cost of universal ART would be between 2% and 5% of GDP[45] and they too would need some outside support. For these eleven countries the total cost of universal ART would amount to US$3.4 billion still affordable by the international community especially with significant input from national governments. All the remaining countries in Africa could afford to pay for universal ART from their own budgets. It is worth noting that in all but two of the remaining 25 countries military expenditure is currently greater than the cost of universal access to ART for all of their citizens.[45]

## Conclusions

In high prevalence settings with generalized epidemics the control of HIV must rely primarily on TasP for the benefit of infected people, to have the greatest impact on the population level incidence of HIV, and to get the greatest impact for the least cost.

Ensuring the availability and accessibility of condoms is important both for individuals but also at a population level and this may be especially important for those that believe themselves to be a risk but are unable to protect themselves in other ways.

Sexually transmitted infections should be treated in their own right as part of a programme of sexual health in the public sector and, for those that are treated as well as for their partners, the reduction in the risk of acquiring HIV will be substantial and very cost-effective.

PreP is more contentious. At an individual level PreP will provide a significant level of protection. However, PreP is likely to be less cost-effective than the other interventions. We have assumed here that PreP would be used daily whereas vaginal microbicides, for example, may only be used before and after a sexual encounter, which would reduce the costs considerably. Especially for women at high risk, including commercial sex-workers, young adolescent women when the incidence of infection is high, and women who believe that they may be at risk from their partners but cannot negotiate condom use with their partners, PreP would be advised.

Behaviour change programmes in themselves are not effective. However, there will always be an important need for psychosocial interventions to support the other interventions, including TasP, in order to deal with stigma and discrimination, to avoid behaviour change disinhibition, to ensure compliance and to mobilize communities. Furthermore, the case of Zimbabwe raises an important question. The incidence of HIV in Zimbabwe began to fall precipitously after 1992[46] for reasons that remain unclear. At that time, ART was not available in Zimbabwe and there is no reason to believe that the natural history of HIV was any different from that in the neighbouring countries. The decline in HIV in Zimbabwe must have been because of changes in behaviour but the problem is to know what those changes were and why they were so significant in Zimbabwe but not in the neighbouring countries, with the possible exception of Tete Province in Mozambique.

This analysis is based primarily on the generalized epidemics in sub-Saharan Africa with a focus on South Africa where $R_0 = 5.8$ and the steady state prevalence in adults is 16%. However, $R_0$ varies among countries[34] making elimination easier in some countries than in others. Furthermore, the steady state or in some cases the peak prevalence of HIV also varies among countries[47] and this will affect the relative costs and benefits. Where the prevalence is very low, as is the case in the Hindi speaking states of north India,[48] the cost of testing all adults would greatly exceed the cost of providing infected adults with ART. In this case contact tracing or provider initiated counselling and testing would be more effective and considerably cheaper than universal routine testing. In countries such as Vietnam where the prevalence varies from about 40% in intra-venous drug users to less than 0.4% in ante-natal clinic women, different strategies may have to be developed for finding and testing people in the different groups.[49]

It is clearly in the best interests of individual people infected with HIV to start ART as soon as possible. This will have the greatest impact on the epidemic and will be the most cost-effective way of controlling the spread of HIV. Any national prevention plan must be built around a programme of early treatment. Other interventions should also be used but in a sensible and appropriate manner. The use of condoms provides good protection and they must be easily and readily available for all those who wish to use them. Sexually transmitted diseases should be treated as a public health problem in their own right and treating sexually transmitted diseases will greatly reduce the risk that individuals face. When people are tested for HIV they should also be tested for other STIs and *vice versa*. PreP offers significant levels of protection and should be made available, wherever possible, to vulnerable people including sex workers and possibly adolescent girls who are sexually active. Male circumcision gives a significant degree of protection for life to individual men and this must be available for all those that want it. The results of rigorous trials of behaviour change programmes have been very disappointing. But even if behavioural and social interventions do not have a direct impact they will play an essential role in persuading people of the benefits of early treatment, in ensuring that those on treatment are fully compliant and that behavioural disinhibition does not compromise the benefits of early treatment. Furthermore the Zimbabwe conundrum remains[46] and suggests that large scale behaviour change is possible and that this can have a dramatic effect on the epidemic.

It is in the best interests of individual people infected with HIV that they start ART as soon as they become infected. A programme of early treatment will lead to the greatest reduction in the prevalence and incidence of ART and offers the only way to eliminate HIV. With good coverage and adherence a programme of early treatment is also the most cost-effective intervention. The primary objective of HIV control programmes in generalized epidemics must be to start as many infected people on ART as soon as possible after they become infected. Other interventions should be used to support this primary intervention. Male circumcision should be available to all young men who want it. Condoms must be readily available for all who need them. PreP should be available to HIV-negative people who are at high risk of infection. Behavioural interventions should focus on ways to mobilize communities around the need for early treatment, counter stigma and discrimination, and ensure high levels of support and adherence for those that are on ART. We have the means with which to eliminate HIV; what remains is the need to mobilize political support and commitment nationally and internationally to finally bring the epidemic to an end.

## Appendix 1. Modelling control

**Dynamic model**

In Appendix 1 we develop a simple dynamical model of HIV infection, in which we assume that everyone starts taking ART either soon after they are infected, under TasP, or just before they die under the other interventions. The model includes people who are susceptible to HIV, infected with HIV, or on ART. We can intervene in two ways, either by reducing transmission or by starting people on ART and reducing the size of the infector pool.

In this model people who would otherwise have died are started on ART. Then we have the following equations for the rate of change of $s$, the proportion of susceptible people (*sensu* Newton),

$$\dot{s} = -\lambda s i + \nu a \qquad 1$$

for the rate of change of $i$, the proportion of infected people, not on ART,

$$\dot{i} = \lambda s i - \alpha i \qquad 2$$

and for the rate of change of $a$, the proportion of people who would have died but are started on ART

$$\dot{a} = \alpha i - \nu a \qquad 3$$

where $\lambda$ is the transmission parameter, $\alpha$ is the rate at which people start ART, and $\nu$ is the rate at which people on ART die. If people do not start ART we set $\nu = 1/\tau$, where $\tau$ is the life expectancy of people infected with ART but who do not start treatment. In this case $a$ is the number of life-years lost because people were not started on ART. We then have

$$R_0 = \frac{\lambda}{\alpha} \qquad 4$$

where $\lambda$ is the transmission parameter, which we can reduce, and $\alpha$ is the rate at which people start ART, which we can increase. Setting the left hand side of Equations 1 to 3 equal to zero at the steady state

$$\alpha \bar{i} = \nu \bar{a} \qquad 5$$

$$\bar{s} = \frac{\alpha}{\lambda} \qquad 6$$

$$\bar{s} + \bar{i} + \bar{a} = 1 \qquad 7$$

so that

$$\frac{\alpha}{\lambda} + \bar{i}\left(1 + \frac{\alpha}{\nu}\right) = 1 \qquad 8$$

$$\bar{i} = \frac{\left(1 - \frac{\alpha}{\lambda}\right)}{\left(1 + \frac{\alpha}{\nu}\right)} = \frac{R_0 - 1}{R_0} \times \frac{\nu}{\nu + \alpha} \qquad 9$$

$$\bar{a} = \frac{\alpha}{\nu}\bar{i} \qquad 10$$

and the incidence of infection $\overline{in}$ is

$$\overline{in} = \lambda \bar{s}\bar{i} = \alpha \bar{i} \qquad 11$$

If $\nu \to \infty$ Equation 9 reduces to the usual expression for a simple susceptible-infected model. For a given value of $R_0$ Equation 9 will overestimate the steady state prevalence because of heterogeneity in sexual behaviour. The simplest way to allow for this is to assume that there are two groups of people; one group is at risk according to Equations 1 to 11, the other group is at no risk. We can then let the prevalence of those at no risk be $\bar{n}$. The measured prevalence before we start the interventions so that no-one is on ART, $P$ must be equal to the prevalence measured as a proportion of all those at risk, infected with HIV and at no risk. Since we want the initial prevalence to be

$$P = \frac{\bar{i}}{\bar{s} + \bar{i} + \bar{n}} \qquad 12$$

we have

$$\bar{n} = \frac{\bar{i}}{P} - \bar{s} - \bar{i} \qquad 13$$

Starting from the initial prevalence, $P$, we calculate the initial incidence,

$$I = \alpha P \qquad 14$$

and the initial number on ART or who are dead but would have been alive without ART

$$A = \frac{\alpha}{\nu}P \qquad 15$$

We now consider the intervention, after it has reached a new steady state, and include those on ART. When we change either the transmission parameter $\lambda$ or the rate at which people start ART the prevalence new steady state prevalence, $\tilde{P}$, is

$$\tilde{P} = \frac{\tilde{i}}{\tilde{s} + \tilde{i} + \tilde{n}} \qquad 16$$

calculating $\tilde{s}$ and $\tilde{i}$ from Equations 6, 9 and 10, but keeping $\tilde{n} = \bar{n}$ since the number of people not at risk should not change. We then calculate $\tilde{A}$ from Equation 15.

## Appendix 2: Note on the costs of ART

Recent studies have estimated the cost of providing ART in PEPFAR programmes in Uganda at US$843 per person per year[50] and of providing ART in Zambia at US$556 per person per year.[51] The price of anti-retroviral drugs continues to fall and the South African government can now purchase fixed-dose combination therapy for less than US$100 per person per month.[52] In a study in Malawi the cost of a rapid HIV test was US$3.50.[53]

## Appendix 3: Cost estimates

In the long term, say 10 years after intervening with a particular programme, the epidemic will reach a new steady state and we calculate the annual cost of maintaining the prevalence at the new steady state and compare this with the reduction in the annual number of new infections between the initial and final steady states. Those people that do get infected with HIV will die if they do not start ART. If they start ART when their CD4$^+$ cell count is very low this will keep them alive but will not affect overall transmission, since those on ART will not infect others. For each intervention we include the cost of maintaining those that are infected with HIV on ART for the rest of their lives.

Since AIDS kills young adults there is a substantial cost to society of letting people die. When a young adult, aged 20 to 30 years dies, society has paid to educate them, shelter them and provide them with health care. The social contract is that they should then work for the next thirty years to pay society back for the money that has been invested in them.[54] For countries in sub-Saharan Africa the median Gross National Product *per capita* is US$1,498 (US$752–US$6,629: 80% range), the median Gross National Income *per capita* is US$645 (US$340 to US$4,220: 80% range).[45] The cost of keeping a person on ART is estimated to be about US$500 per year so that if we let the cost of an AIDS death equal the cost of keeping that person alive on ART the calculations will be conservative as regards cost in almost all of sub-Saharan Africa. We therefore assume that everyone starts ART either early under TasP or just before they would have died under other interventions.

## Appendix 4: Cost-benefit calculations

**Treatment as Prevention**

Let the cost of first finding and then keeping one person on ART for one year be $C_{ART}$. If the prevalence of infection is $P_{HIV}$ then it will take, on average, $1/P_{HIV}$ tests per year to find one person infected with HIV and the cost will be $C_{Test}(1-P_{HIV})/P_{HIV}$. The cost of putting all HIV-positive people on ART for one year is

$$C_{TasP} = C_{ART} P_{HIV} + C_{Test}(1 - P_{HIV})/T_{Test} \quad 17$$

where $T_{Test}$ is the test interval in years and Equation 17 holds when estimating the individual and the short-term impact.

In the short term the cost of testing will exceed the cost of keeping a person on ART for one year when

$$P_{HIV} = \frac{C_{Test}/T_{Test}}{C_{ART} + C_{Test}/T_{Test}} \quad 18$$

if the cost of a test is $5, the cost of keeping a person on ART for one year is $500, and testing is done once a year, the cost of testing will exceed the cost of ART for one year when the prevalence is less than 1%. However, if testing were done monthly, say, the cost of testing will exceed the cost of ART when the prevalence is less than 12%. When we consider the population level effect we allow for the coverage and compliance assumptions in Figure 3B

In the long-term the number of people on ART will be, equal to the prevalence of HIV times the rate at which people are tested for HIV divided by the rate at which people on ART die (Equation 15, Appendix 1). For the model in Appendix 1 this will be equal to the life expectancy on ART divided by the test interval so that

$$C_{TasP} = C_{ART} P^*_{HIV} \left( L_{ART}/T_{Test} \right) \\ + C_{Test}(1 - P^*_{HIV})/T_{Test} \quad 19$$

where we calculate the long-term steady state prevalence, $P^*_{HIV}$, using Equation 16 in Appendix 1. In the long term the cost of testing will exceed the cost of keeping a person on ART for one year when

$$P^*_{HIV} = \frac{C_{Test}/T_{Test}}{C_{ART} L_{ART} + C_{Test}/T_{Test}} \quad 20$$

and if the life expectancy on ART is 30 years then the critical prevalence will be 0.03%.

### Condom promotion

The cost incurred if a person uses condoms regularly over a period of one year is the cost of a condom, $C_{Cond}$, times the number of sexual encounters that person has in one year, $N_{SE}$, so that

$$C_{CP} = C_{Cond} N_{SE} \quad 21$$

Here we assume that people at risk of HIV have sex on average 100 times per year.

### Pre-exposure prophylaxis

Under PreP we calculate the cost of keeping one HIV-negative person on prophylaxis for one year so that

$$C_{PreP} = C_{Drugs} + C_{Test}/T_{PreP} \quad 22$$

where $C_{Drugs}$ is the cost of the drugs for one year, $C_{Test}$ is the cost of testing a person for HIV and $T_{PreP}$ is the time between tests for a person on PreP to confirm that they have not yet been infected with HIV.

For PreP the cost of testing would exceed the cost of the drugs if the time between tests were less than

$$T_{Prep} = \frac{C_{Test}}{C_{PreP}} \quad 23$$

If the cost of PreP is about US$1 per day, or US$365 per year, and the cost of a test is about US$10 then the cost of testing would exceed the cost of drugs if testing were done more often than about 36 times per year.

### Male circumcision

Male circumcision is different from the other interventions because it incurs a once-off cost which gives a degree of protection for life. For the individual cost and the short term cost we consider the once-off cost of MC. For the long term cost we spread the cost of one circumcision over the time for which a person is at risk of HIV, $T_{Risk}$, so that the annual cost of MC is, in the long term,

$$C_{MC} = C_{Circ}/T_{Risk} \quad 24$$

and we assume that $T_{Risk}$ is ten years.

### Treating sexually transmitted infections

The cost of treating a case of a sexually transmitted disease is more difficult to estimate. We need to know the incidence of STIs, $I_{STI}$, as this will determine the number of people who will present with an STI each year. The cost of treating all STIs is then

$$C_{STI} = C_{Treat} I_{STI} \quad 25$$

where $C_{Treat}$ is the cost of treating one STI. There is evidence that syndromic management of STIs reduces the incidence of HIV but it should be noted that in the one trial that showed a significant reduction in the incidence of HIV the prevalence of all STIs increased over the course of the study in both the intervention and control arms.[23]

### Behaviour change

For behaviour change interventions estimate $C_{BC}$ the cost per person covered of running a behaviour change programme for one year.